\documentclass[%
 reprint,pre,
showpacs,
 amsmath,amssymb,
 superscriptaddress,
 aps]{revtex4-1}

\usepackage{amsmath}
\usepackage{amssymb}
\usepackage{graphicx}
\usepackage{dcolumn}
\usepackage{bm}
\usepackage{hyperref}
\usepackage{verbatim}
\usepackage{color}
\usepackage{xcolor}
\usepackage[version=4]{mhchem}
\usepackage{appendix}
\usepackage{physics}

\usepackage{verbatim}
\usepackage{graphics}
\usepackage{longtable} 
\usepackage{subfigure}
\usepackage{amsmath}
\usepackage{wrapfig}
\usepackage{epsfig}
\usepackage{float}
\usepackage{graphicx}
\usepackage{array}
\usepackage{psfrag}
\usepackage{color}
\usepackage{bbold}
\usepackage{enumerate}
\usepackage{layouts}

\definecolor{azure}{rgb}{0.0, 0.5, 1.0}
\definecolor{asparagus}{rgb}{0.53, 0.66, 0.42}
\definecolor{ballblue}{rgb}{0.13, 0.67, 0.8}
\definecolor{sgreen}{rgb}{0.0, 0.8, 0.35}
\definecolor{darkgreen}{rgb}{0.0, 0.5, 0.0}
\definecolor{sred}{rgb}{0.9, 0.6, 0.4}
\usepackage{textcomp}

\usepackage{tikz}

\usepackage[]{color}

\begin{document}

\title{
Metastability and ripening  of multi-component liquid mixtures 
}

\author{Giacomo Bartolucci}
\email{bartolucci@ub.edu}
\affiliation{Department of Condensed Matter Physics, Universitat de Barcelona, 08007  Barcelona, Spain}
\affiliation{Universitat de Barcelona Institute of Complex Systems (UBICS), Universitat de
Barcelona, 08028 Barcelona, Spain}
\author{Fabrizio Olmeda}
\email{fabrizio.olmeda@ista.ac.at}
\affiliation{Institute of Science and Technology Austria, Am Campus 1, 3400 Klosterneuberg, Austria}

\begin{abstract}
Understanding how multi-component liquid mixtures undergo phase separation is central to elucidating biophysical organization in the cell. Here, combining analytical and numerical results, we characterise the dynamics of mixtures with disordered interactions among the components. 
We first study how two coexisting phases become unstable, leading to multiphase coexistence. We then show that the scaling of droplet radius as $t^{1/3}$ and droplet number as $n^{-2/3}$, characteristic of Ostwald ripening in two dimensions, 
can be severely delayed. 
This delay arises from glass-like relaxation and the emergence of long-lived metastable states characterized by different wetting angles. 
\end{abstract}

\maketitle

\textit{Introduction.--- }
Phase separation plays a crucial role in soft matter with applications ranging from cell biology~\cite{brangwynneGermlineGranulesAre2009,hymanLiquidLiquidPhaseSeparation2014, bananiBiomolecularCondensatesOrganizers2017}, to origin of life~\cite{oparinOriginLife1952,haldaneOriginLife1929,bartolucciSelectionPrebioticOligonucleotides2022}, and food science~\cite{mathijssenCulinaryFluidMechanics2023, bartolucci2025phase}. Crucially, the majority of experimentally relevant mixtures undergoing phase separation comprise a large number of components. For example, in the cell cytoplasm, the formation of membraneless organelles via phase separation is orchestrated by thousands of proteins~\cite{vanderleeClassificationIntrinsicallyDisordered2014,youPhaSepDB30Comprehensive2025}. 

Despite substantial progress in elucidating the thermodynamics of multicomponent mixtures~\cite{shrinivasPhaseSeparationFluids2021,zwickerEvolvedInteractionsStabilize2022}, mostly focusing on characterizing spinodal instabilities \cite{searInstabilitiesComplexMixtures2003,jacobsPhaseTransitionsBiological2017, carugnoInstabilitiesComplexFluids2022, thewes2023composition}, their out-of-equilibrium dynamics remain mostly underexplored \cite{shrinivasMultiphaseCoexistenceCapacity2022,thewes2024mobility, brazteixeiraLiquidHopfieldModel2024}, particularly regarding droplet ripening and metastability. While linear spinodal analysis captures the initial growth of fluctuations from a homogeneous well-mixed state, the long-time evolution is governed by the approach to the binodal (coexistence) manifold, where nonlinearity becomes important. In binary mixtures, late-stage coarsening follows the classical Ostwald ripening scenario, for which the characteristic droplet size grows in time as $t^{1/3}$ \cite{LifshitzSlyozov1961,Wagner1961}. 
Ripening in mixtures of a few components has been studied extensively, even under sustained non-equilibrium fluxes \cite{zwickerGrowthDivisionActive2017, bauermannTheoryReversedRipening2025}. Although the prevailing intuition has suggested that multi-component mixtures should inherit the binary scaling \cite{bray2002theory}, numerical studies are still elusive. 

\begin{figure}[t] 
\includegraphics[width=1\linewidth]{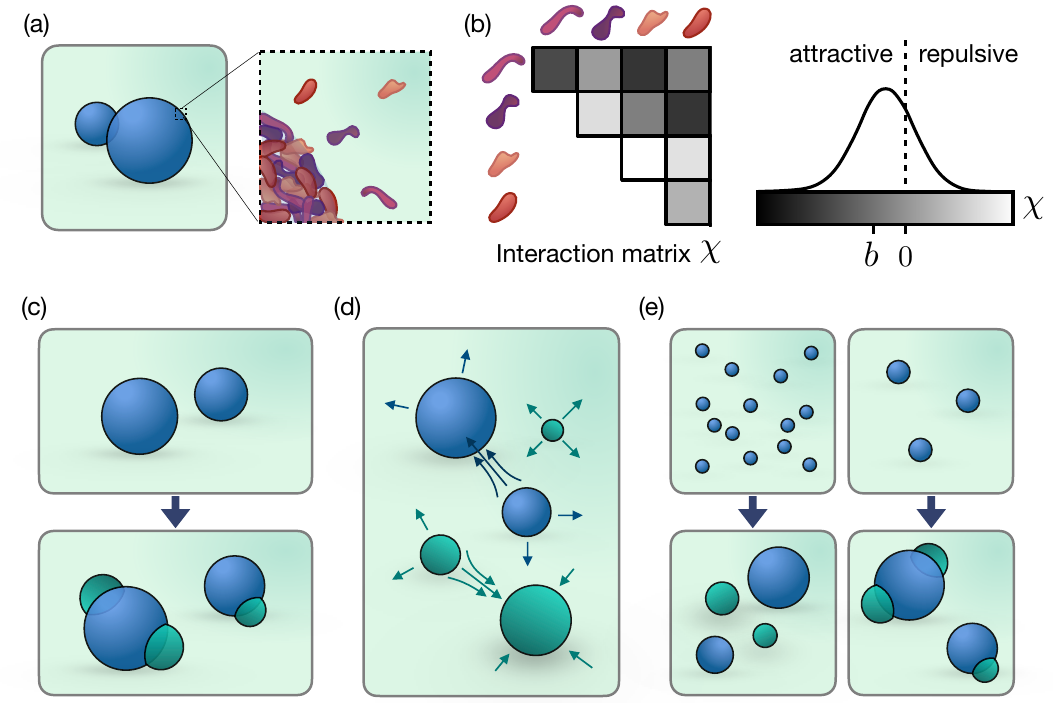}
\caption{Schematic for the dynamics of multi-component phase-separating mixtures. (\textbf{a}) Two coexisting phases characterised by different solute compositions (inset) (\textbf{b}) pairwise solute interactions are encoded in a matrix $\chi$ composed of Gaussian random variables with mean $b$. (\textbf{c})  Two phases can be locally stable, while three phases emerge at later times. (\textbf{d}) Ripening dynamics in the presence of multiple coexisting phases. (\textbf{e}) Different initial conditions can lead to different long-lived metastable states charcaterized by different wetting angles.
}
\label{fig:gra_abs}
\end{figure}

Here, apply the formalism of soft spin glasses to study phase separation in multi-component mixtures with random interactions, see Fig.s ~\ref{fig:gra_abs}(a) and (b). Combining theory with large-scale numerical simulations, we delineate and characterise three distinct dynamical regimes of phase separation.
We first revisit the early spinodal decomposition, and then characterise the onset of linear instabilities within two locally stable phases that lead to multiphase coexistence, Fig.~\ref{fig:gra_abs} (c). We finally address ripening in the presence of multiple distinct phases Fig.~\ref{fig:gra_abs} (d), showing that as the number of solutes increases, ripening is slown down, leading to deviations from Ostwald scaling. We finish demonstrating that different initial conditions cause the mixture to remain trapped in different long-lived metastable states, which is a hallmark of glass-like dynamics, Fig.~\ref{fig:gra_abs} (e).

\textit{Field-theory of multi-component mixtures.--- }We describe the multi-component mixture in the $T$-$V$-$N_i$-ensemble, $N_i$ being the particle number of the component $i=1,\dots,M+1$. 
In this work, we focus on incompressible mixtures in which every component has identical molecular volume $v$. We proceed by introducing the volume fractions of each component $\phi_i = N_i v/V$. Volume conservation implies that
\begin{align}
\label{eq:phi_t_inco}
	\phi_{M+1} = 1 - \phi_\text{tot}\,,  \qquad \phi_\text{tot} =  \sum_{i=1}^{M} \phi_i \,.
\end{align} 
We denote the $M+1$ component as the solvent and the components $i=1,\dots,M$ as the solutes. We then write the field theory describing the spatio-temporal evolution of each solute volume fraction field,
\begin{equation}
\label{eq:d_phi_i_dt}
    \partial_t \phi_i = \div \Big\{ \Gamma(\boldsymbol{\phi})\grad \left[\frac{\partial F}{\partial \phi_i} +  \eta_i\right] \Big\}\,
\end{equation}
where $\Gamma(\boldsymbol{\phi})$ is a mobility coefficient and $\eta$ is a Gaussian white noise that satisfies the fluctuations dissipation theorem. We also introduced $F$, the Helmholtz free energy, which can be written in terms of the free energy density,
\begin{align}
\label{eq:F_helm}
    F =  \int \dd V \left[ f\left( T,V,\{\phi_i\} \right) + \sum_{i=1}^{M}  \frac{\kappa_i}{2} \left( \nabla\phi_i \right)^2 \right].
\end{align}

In this work, we consider the Flory-Huggins free energy density, which for multi-component mixtures are ~\cite{floryThermodynamicsHighPolymer1942,hugginsThermodynamicPropertiesSolutions1942},
\begin{align}
\label{eq:f}
    f = \frac{k_B T}{v} \left[ \;\sum_{i=1}^{M+1} \phi_i \ln \phi_i + \sum_{i, j=1}^{M} \frac{ \chi_{ij} }{ 2 k_B T} \, \phi_i  \phi_j  \right],
\end{align}

Here, $\chi_{ij}$ is a symmetric matrix that parametrizes the interaction between solute $i$ and $j$. In this work, we sample the entries of $\chi_{ij}$ from a gaussian distribution with mean $b$ and standard deviation $\sigma \sqrt{M}$ \cite{thewes2023composition,searInstabilitiesComplexMixtures2003}. The scaling of the variance as $\sqrt{M}$ leads to a bound free energy density as the number of solutes diverges, while $\phi_\text{tot}$ is kept constant, i.e. $\phi_i \sim \mathcal{O}(1/M)$. With this scaling, all the thermodynamic quantities, such as the spinodal curve, become independent of $M$, the number of solutes (see \cite{SupplementaryInformation} and following).
Here, by applying methods from spatially extended systems composed of many interacting components, \cite{olmeda2023long,baron2020dispersal}, we can simplify the previous equation, passing from $M$ coupled equations, to $M$ uncoupled equations:  

\begin{equation}
\begin{split}
\label{eq:MSRJD}
& \partial_t \phi_i = \div \big\{ \Gamma(\phi_i) \grad \big[\log(\phi_i) - \log(1-\phi_{\mathrm{tot}}) + b \phi_\text{tot}  \\
&+ M \sigma^2 \int \int dx' dt' G_i \phi_i + \kappa \grad^2\phi_i + W_i + \eta_i \big] \big\}\,,
\end{split}
\end{equation}

where $G_i(\Vec{x},\Vec{x}',t,t')$ is a response function and in the integrals, we simplify the notation for dummy indexes.  The noise $W_i$ has variance:  $\langle W_i(\vec{x},t) W_j(\Vec{x}',t') \rangle = \delta_{ij}\delta_{t-t'} \delta(\vec{x}' - \vec{x}) C_{i,j}(\vec{x},\Vec{x}',t,t')$. The quantity $C_{ij}$ inside the noise variance is the  spatio-temporal correlation  between fields. 
Below, we will show the applicability of this framework to the spinodal instability, benchmarking some known results of random matrix theory. Moreover, the formulation in Eq.~\eqref{eq:MSRJD} allows to easily include higher-order solute interactions and can benefit from methods developed for soft-spin glasses \cite{crisanti1993spherical,cugliandolo2023recent}.


\textit{Early stage dynamics: Spinodal.--- }
The dynamics following a perturbation of a well-mixed mixture can be studied by performing an expansion around the homogeneous concentration values: $\phi_i(x) = \bar{\phi}_i+\psi_i(x)$. For simplicity, we consider the case where all the solutes are present in equal amounts, $\bar{\phi}_i = \bar{\phi}_\text{tot}/M\,\, \forall i$ and we neglect noise. The stability analysis of the Hessian associated to the free energy in Eq.~\eqref{eq:MSRJD} 
leads to a parametric curve, namely the spinodal, below which infinitesimal perturbations grow, ultimately leading to phase separation. To linear order in Fourier space, the stability of Eq.~\eqref{eq:MSRJD} is given by,

\begin{equation}
\partial_t \psi_i(t) = -\Omega_i\left[\, \delta_i\,\psi
+r + \sigma^2 M \int dt'G_i \psi_i \,\right],
\end{equation}

where $\delta_i = 1/ \bar{\phi}_i - \kappa k^2$ and $\Omega = \Gamma(\bar{\phi})k^2$ and $r = \sum_i \psi_i(b+(1-\phi_{\mathrm{tot}})^{-1})$, a rank-1 perturbation. We now consider two different scenarios: condensation and demixing. The latter is defined as the instability along the transverse direction, i.e. $\sum_{i} \psi_i = 0$, which corresponds to the scenario in which the components ``shuffle'' without being locally enriched or depleted.  In this case $r =0$  and we find that the response function $G$  satisfies the quadratic equation (invoking time-translation invariance and dropping the index),

\begin{equation}
\label{eq:resolvent_response}
\Omega\, M\sigma^2 \tilde{G}(s)^2 +(\Omega \delta + s)\,\tilde{G}(s) + \Omega = 0 \, ,
\end{equation}

where we took the Laplace transform of the response functions in time. 
Response functions satisfy the same equations as the resolvent of a Gaussian symmetric random matrix \cite{Wigner1958}, thus the response $\tilde{G}(s)$ is the resolvent $G(z)$ evaluated at $z = -s/\Omega$. The spinodal is then determined by the condition that the resolvent develops a branch point at $s=0$, corresponding to the lower edge of the eigenvalue spectrum touching zero. This yields, for large $\phi_{\mathrm{tot}}$,
\begin{equation}
k_B T_{\mathrm{spin}} = 2\,\sigma \,\phi_{\mathrm{tot}}, \quad \phi_\text{tot} > \phi_\text{tot}^*\,
\end{equation}
where $\phi_\text{tot}^*$ has not yet been determined and corresponds to the transition value between condensation and demixing. Condensation is determined instead by a perturbation along the longitudinal direction, i.e. $\sum_i \psi_i \neq 0$. Physically, this corresponds to the case in which all solutes are enriched or depleted in one phase. In \cite{SupplementaryInformation} we show that following similar  steps we can analytically derive the spinodal for condensation-type instability and altogether,
\begin{align}
    \label{eq:spin}
    k_BT_{\mathrm{spin}} =  
    \begin{cases} 
    - b \; \phi_\text{tot}(1-\phi_\text{tot}) \qquad &\phi_\text{tot} < \phi_\text{tot}^* \\
    2 \sigma \,\phi_\text{tot} \qquad &\phi_\text{tot} > \phi_\text{tot}^*
    \end{cases}
\end{align}

where $\phi^* = 1+2\sigma/b$. As mentioned above, with the standard deviation of our random matrix scaling like $\sqrt{M} \sigma$, as in~\cite{thewes2023composition}, the spinodal temperature is independent of the number of solutes $M$, see also~\cite{SupplementaryInformation}. 
As the temperature approaches $T_{\mathrm{spin}}$, the left edge of the eigenvalue spectrum of the random relaxation operator approaches zero, leading to critical slowing down. The mean-square perturbation decays in the demixing regime as
$|\psi(t)|^{2} \sim \int d\lambda\, \rho(\lambda)\, e^{-2\lambda t},$
where $\rho(\lambda)$ denotes the spectral density of relaxation rates. For a Wigner semicircle density, $\rho(\lambda)$ is finite at $\lambda=0$, and the long-time asymptotics is $|\psi(t)|^{2} \sim t^{-1}$.  

\begin{figure}[t] 
\includegraphics[width=1\linewidth]{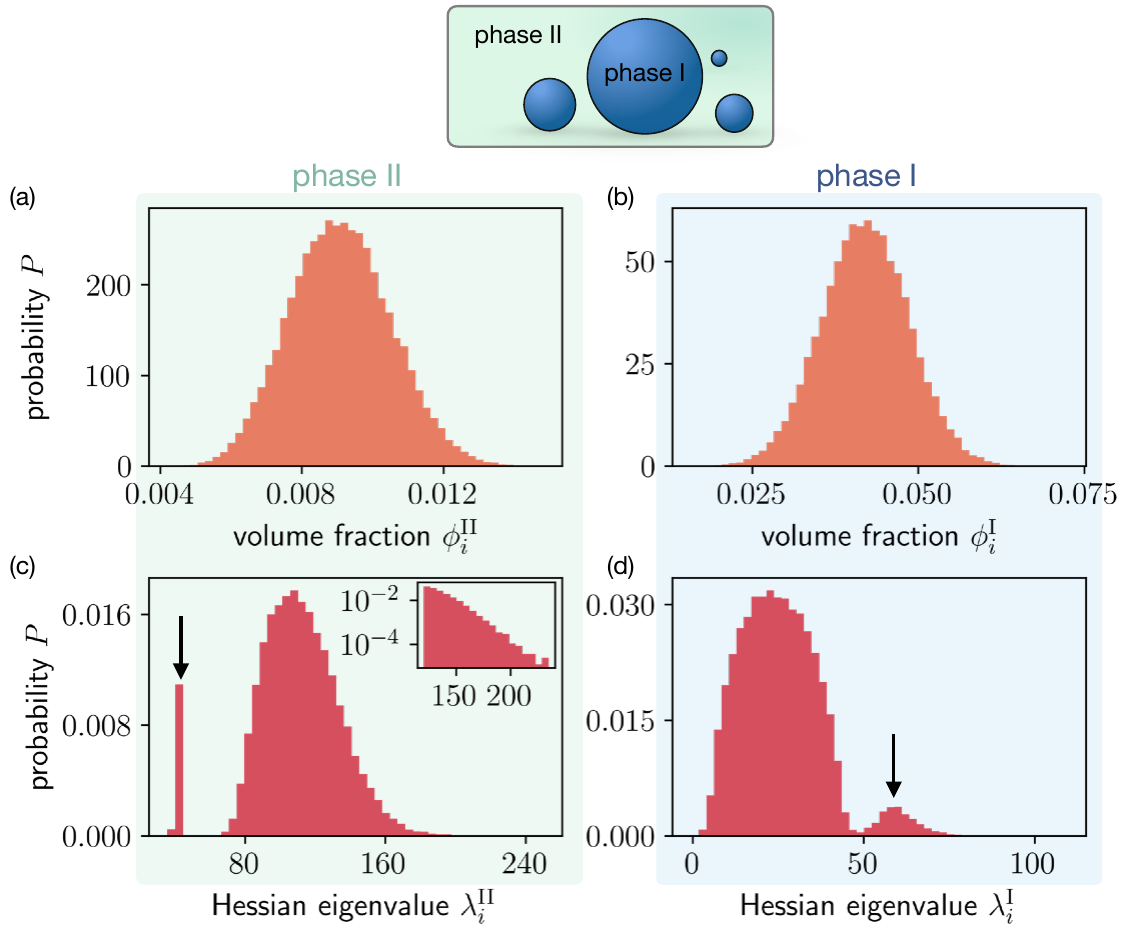}
\caption{Top: schematic of the dilute (phase II) and dense phase (phase I). (\textbf{a}-\textbf{b}) Numerical distribution of solute volume fraction in the dilute (II) and dense (I) phases. (\textbf{c}-\textbf{d})
Eigenvalues of the Hessian matrix evaluated at the dilute (II) and dense (I) phases, respectively. The peak corresponding to isolated eigenvalues lies to the left of the continuum spectrum in phase II, and to the left of it in phase I. The continuum part can deviate from the Wigner semicircle developing an exponential tail (inset).}
    \label{fig:ran_mat}
\end{figure}

\textit{Intermediate stage dynamics: Onset of multi-phase coexistence.--- }
After the initial transient in which perturbations evolve following the eigenvectors of the Hessian matrix \cite{carugnoInstabilitiesComplexFluids2022}, the system demixes into two or multiple phases (more on it later), whose volume fractions lie on the binodal manifold and can be found via the Boltzmann construction ~\cite{weberPhysicsActiveEmulsions2019}. 
We study the intermediate stage where the system has demixed into two locally stable phases such that $\bar{\phi}_i = V^\text{I} \phi^\text{I}_i + V^\text{II} \phi^\text{II}_i$. We identify phase II and I with the $\phi_\text{tot}$-poor and $\phi_\text{tot}$-rich phases, respectively, $V^\text{I}$ and $V^\text{II}=V-V^\text{I}$ being their volumes. We characterize the dynamics of perturbations in both phases: $\phi(x) = \phi^\alpha_i + \psi^\alpha(x)$, where $\alpha =$ I,II. Similarly to the spinodal analysis, the intermediate-stage dynamics is governed by the Hessian matrix evaluated in the two coexisting phases: $H^\alpha = H(\phi^\alpha)$. Again, we can recast the equation for the perturbations in this regime as,
\begin{equation}
\label{eq:intermidiate_dynamics}
\partial_t \psi^\alpha(t) = -\Omega\left[\, \delta^\alpha \,\psi^\alpha + r
+ M \sigma^2 \int dt' G_i \psi^{\alpha}_i \,\right],
\end{equation}
where $\delta^\alpha  = 1/ \phi^\alpha_i - \kappa k^2$ and $ \phi^{\alpha}_i$ is the composition in one of the phases of the component $i$. Note that, even when the total volume fractions are equal for all the components, i.e. $\bar{\phi}_i = \phi_\text{tot}/M$, the volume fractions $\phi^{\alpha}_i$ in each phase might be different for each component. In order to gain insight on this intermediate regime, we numerically found the coexisting phases via the Maxwell construction and estimated the distribution of the volume fractions in each phase, $\phi_i^\text{II}$ and $\phi_i^\text{I}$, see Fig.~\ref{fig:ran_mat}, (a), (b).
In Fig.~\ref{fig:ran_mat}, (b), (c), we show the corresponding eigenvalues of the Hessian matrix, i.e. $\lambda_i^\text{II}$ and $\lambda_i^\text{I}$.

The key difference with respect to the spinodal instability is that the resolvent and  response functions deviate from the superposition of the Wigner semicircle and a lone eigenvalue. In \cite{SupplementaryInformation} we give details of the exact solution of the resolvent and the eigenvalue distribution. Writing the solutions of Eq.\eqref{eq:intermidiate_dynamics} upon projection on the eigenvectors of the hessian computed at the coexisting phases, each relaxation time-scale of the individual modes is proportional to the eigenvalues: $ \tau_i = 1/ \lambda_i$.
In this intermediate regime, there can be fast equilibration as many time-scales are small, 
Fig.~\ref{fig:ran_mat} (a) inset in the left panel. At the same time, when the left edge of the binodal approaches zero, we expect a similar behaviour as for the spinodal. As incompressible mixtures of $M$ solutes might demix into a maximum of $M+2$ distinct phases~\cite{gibbsEquilibriumHeterogeneousSubstances1879}, the previous reasoning can be extended to each of the phases, leading to a rich dynamical behaviour. Together, this signals that a system might be trapped for a long-time in minima of the free-energy characterized by two or more phases. Furthermore, locally in each phase, the systems might experience different relaxation dynamics, having a fundamental impact on the coarsening.

\textit{Late stage dynamics: Ripening.--- } 
In order to investigate the coarsening of multi-component mixtures, we follow similar steps as in binary mixtures \cite{weberPhysicsActiveEmulsions2019}. To this end, we consider that the system is initially phase separated into multiple droplets with compositions given by the binodal. In particular, as the equilibrium concentrations $\mathbf{\phi}_i^\alpha$ are known, in the mean-field limit the radius of a droplet of a phase $\beta$ surrounded by a ``sea" of phase $\alpha$ evolves according to \cite{hoyt1998coarsening,philippe2013ostwald},
\begin{equation}
\begin{split}
\label{eq:dynamics_droplet_phase}
& \partial_tR_i^{\beta} = \frac{2 \gamma }{  R_{\beta} \, \Delta  \bar{\pmb{\phi}^T}^{\alpha \beta} D^{-1}\, \Delta  \bar{\pmb{\phi}}^{\alpha \beta}} \left[
\frac{\Delta  \bar{\pmb{\phi}^T}^{\alpha \beta} H^\alpha \Delta  \pmb{\phi}^{\alpha}_{\infty}}{2\gamma}
- \frac{1}{R_i^{\beta}}\right] \, , \\
&V \bar{\phi}_i  = \sum_i \frac{4\pi}{3} R^{\beta \,  ^{\large 3}}_{i} \, \phi_i^{\beta} + \phi^{\alpha}_{i, \infty}(V - \sum_i \frac{4\pi}{3} R^{\beta \,  ^{\large 3}}_{i}) \, ,    
\end{split}
\end{equation}

 where $\Delta  \bar{\pmb{\phi}}^{\alpha \beta} = \pmb{\phi}^{\beta} - \pmb{\phi}^{\alpha}$ and  $ \Delta \pmb{\phi}^{\alpha}_{\infty} =  \pmb{\phi}_{\infty}^{\alpha} - \pmb{\phi}^{\alpha}$  and the volume fractions  $\pmb{\phi}_{\infty}^{\alpha}$ at infinity are given by volume and incompressibility constraints. Every quantity, such as the mobilities and the hessian, is evaluated with the volume of the outside phase $\alpha$ given by the Maxwell construction. Numerically solving Eq.s~\eqref{eq:dynamics_droplet_phase}, leads to Ostwald ripening, i.e. mean droplet area and droplet number scaling like $\langle A \rangle \sim t^{2/3}$ and $n \sim t^{-2/3}$ respectively, irrespectively of the number of components, as also claimed in 
 ~\cite{bray2002theory,shrinivasPhaseSeparationFluids2021,kim2018ostwald,hoyt1998coarsening}

\begin{figure}[t] 
\includegraphics[width=1\linewidth]{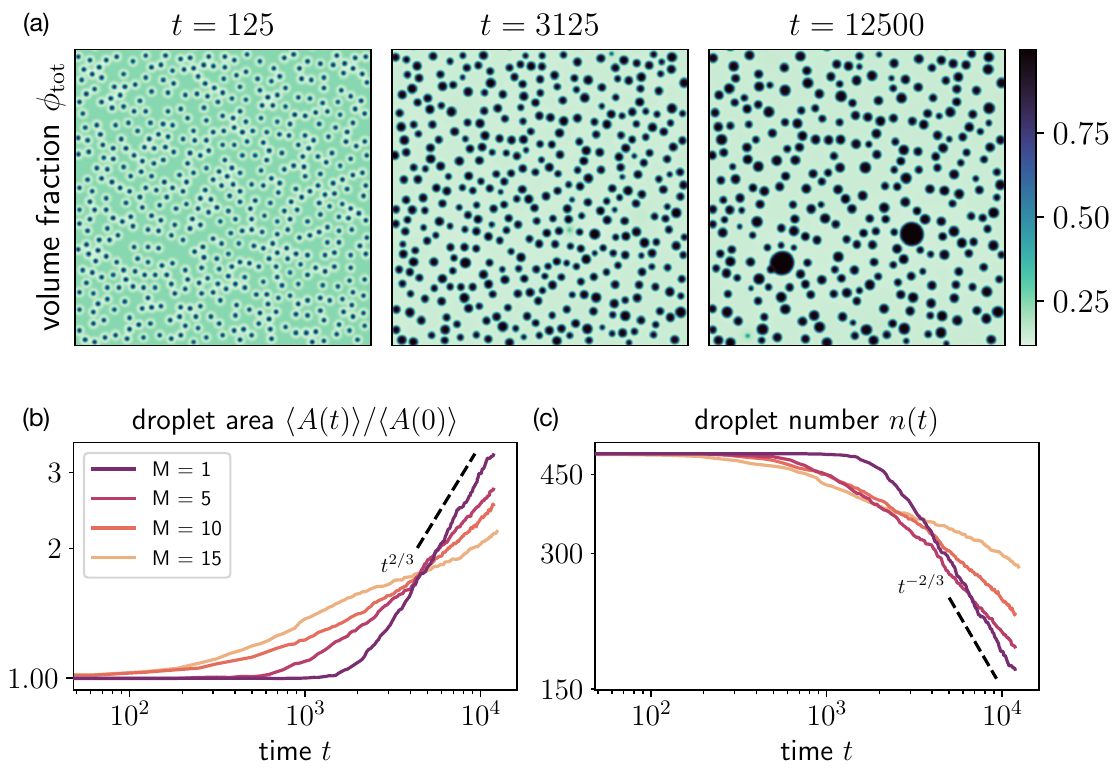}
\caption{(\textbf{a}) Snapshots of the total solute volume fraction, $\phi_\text{tot}$ showing ripening of a mixture of $M=15$ solutes. (\textbf{b}) average radius and (\textbf{c}) number of $\phi_\text{tot}$-rich droplets as a function of time, for $M=1$, $M=5$, $M=10$, and $M=15$ number of solutes. Ostwald ripening scaling is highlighted with a dashed line.}
\label{fig:ripe_all}
\end{figure} 


Here, via numerical simulations of the field theory \eqref{eq:d_phi_i_dt}, we found deviations from these power laws during the initial ripening stage. 
Specifically, we chose equal homogeneous volume fractions for the solutes $\bar{\phi}_i=\bar{\phi}_\text{tot}/M$ and we locate the spinodal for a given realization of the random matrix $\chi_{ij}$. We then study the dynamics at the spinodal temperature, introducing initially a fixed number of pointlike perturbations, which will act as seeds for the droplets. 
In Fig.~\ref{fig:ripe_all} (a), we show the dynamics of the total volume fraction field for $\phi_\text{tot}=0.3$ and the solute number $M=15$. The initial $N=500$ perturbations grow into droplets, which then start to ripen, see also Movie 1 in~\cite{SupplementaryInformation}. With a simple threshold, we can compute the number of $\phi_\text{tot}$-rich droplets $n(t)$ and their mean area $A(t)$, as a function of time. In Fig.~\ref{fig:ripe_all} (b-c) in \cite{SupplementaryInformation}, we display the results comparing binary mixtures ($M=1$) with multicomponent mixtures ($M=5,10,15$).
Interestingly, multi-component mixtures begin to coarsen before the binary case. However, their initial ripening is slower and deviates from the Ostwald regime, displayed with black dashed lines in Fig.~\ref{fig:ripe_all} (b-c). The regime of Ostwald ripening is eventually recovered only after a long transient, in which about half of the initial droplets have already dissolved. For a comparison between the binary and the multicomponent cases see also Movies 1-4 in~\cite{SupplementaryInformation}. 

Note that, in multicomponent mixtures, multiple phases can form, and having considered $\phi_\text{tot}$-rich droplets corresponds to summing over all the distinct phases. In the next section, we study the ripening dynamics of each phase separately. We expect at least one phase to deviate from Ostwald ripening, since we have shown that their sum does.

\begin{figure}[b] 
\includegraphics[width=1\linewidth]{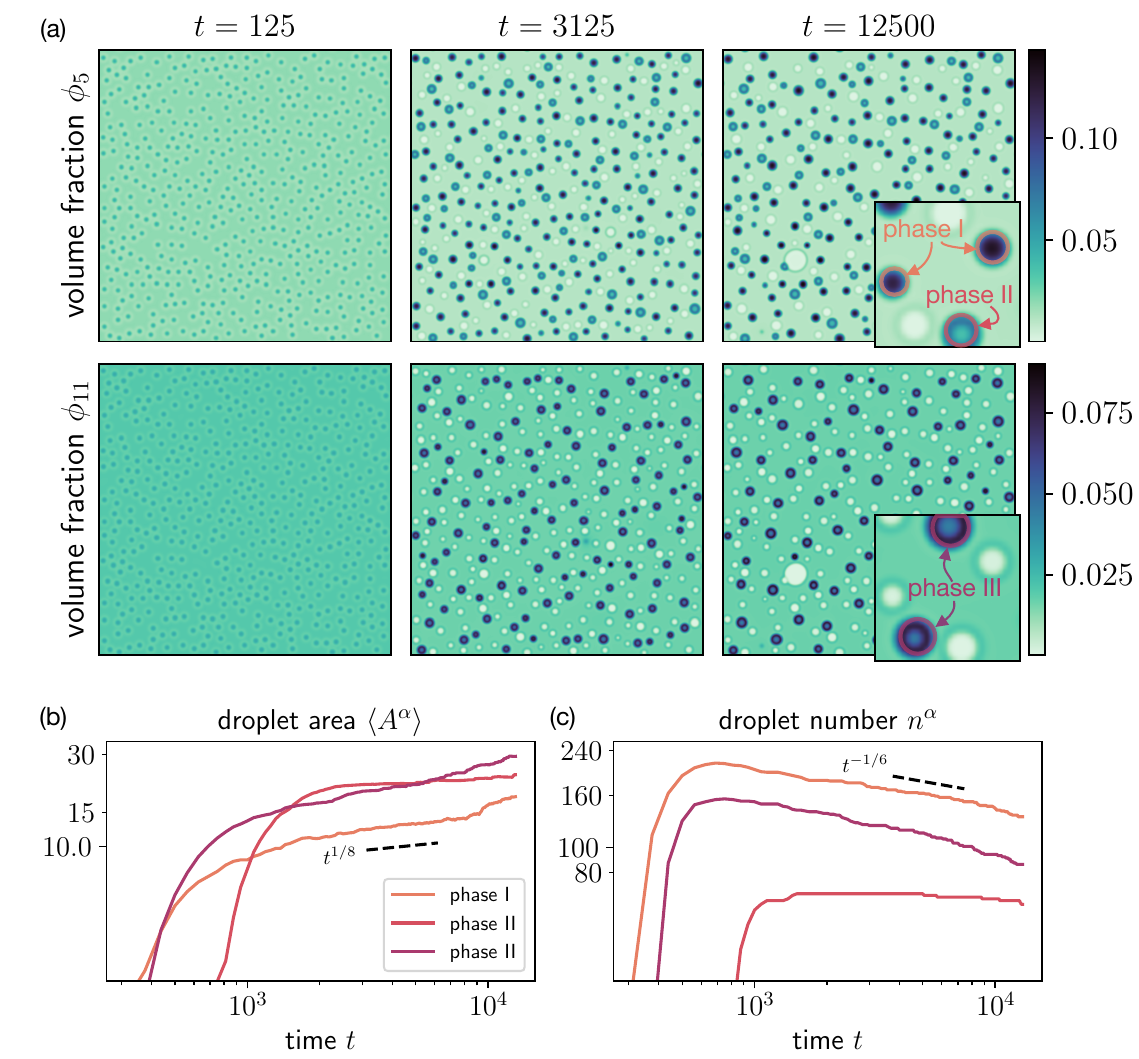}
\caption{(\textbf{a}) Snapshots two solute volume fractions, $\phi_5$ and $\phi_{11}$ showing ripening of a mixture of $M=15$ solutes. (\textbf{b}) average area $A^\alpha$ and (\textbf{c}) number of droplets $n^\alpha$ as a function of time, for two distinct phases $\alpha = \text{I},\text{II}$. Dashed black lines show power laws with exponents significantly different from Ostwald ripening.}
    \label{fig:ripe_comps}
\end{figure}

\textit{Multiphase ripening dynamics.--- }
As mentioned above, incompressible mixtures with $M$ solutes might demix into a maximum of $M+2$ distinct phases~\cite{gibbsEquilibriumHeterogeneousSubstances1879}. Indeed, inspecting volume fractions of single components, in the example with $M=15$ presented in Fig.~\ref{fig:ripe_comps} (a), we can identify 4 distinct droplet phases immersed in a solvent-rich phase, see also~\cite{SupplementaryInformation}. 

In Fig.~\ref{fig:ripe_comps} (a), we show the dynamics of two volume fraction fields, i.e. $\phi_5$ and $\phi_{11}$, of the same simulation presented above. The $N=500$ droplets initially grow into small droplets rich in all the solutes. After this short transient, droplets differentiate into distinct phases, each one enriched with specific solutes, see also Movie 1 in \cite{SupplementaryInformation}. With appropriate thresholds, we can compute the number of droplets of each phase, $n^\alpha$ and the corresponding mean area $A^\alpha$, as a function of time. In Fig.~\ref{fig:ripe_all} (b-c), we display the results comparing binary mixtures ($M=1$) with multicomponent mixtures ($M=5,15$).
Interestingly, distinct phases deviates from the Ostwald regime, see power laws displayed with black dashed lines in Fig.~\ref{fig:ripe_comps} (b-c). This confirms that Ostwald ripening might be recovered after long transients, beyond the time scales achieved with our numerical analysis. 

We speculate that deviations from the power law behaviour characteristic of Ostwald ripening arise due to spatial heterogeneities. Indeed, the crucial assumption underlying Ostwald ripening, as well as Eq.~\eqref{eq:dynamics_droplet_phase}, is that each droplet can be considered isolated, and  ``feels'' the same volume fractions outside its interface, in the far field. From numerical simulations, especially in the presence of multiple phases, this is clearly not observed, see Fig.~\ref{fig:ripe_all} and Movie 1 in \cite{SupplementaryInformation}. Moreover, is already known, that correction due to spatial anisotropy may lead to a change of the scaling of the droplet distribution~\cite{che1995spatial}.

\begin{figure}[!ht] 
\includegraphics[width=1\linewidth]{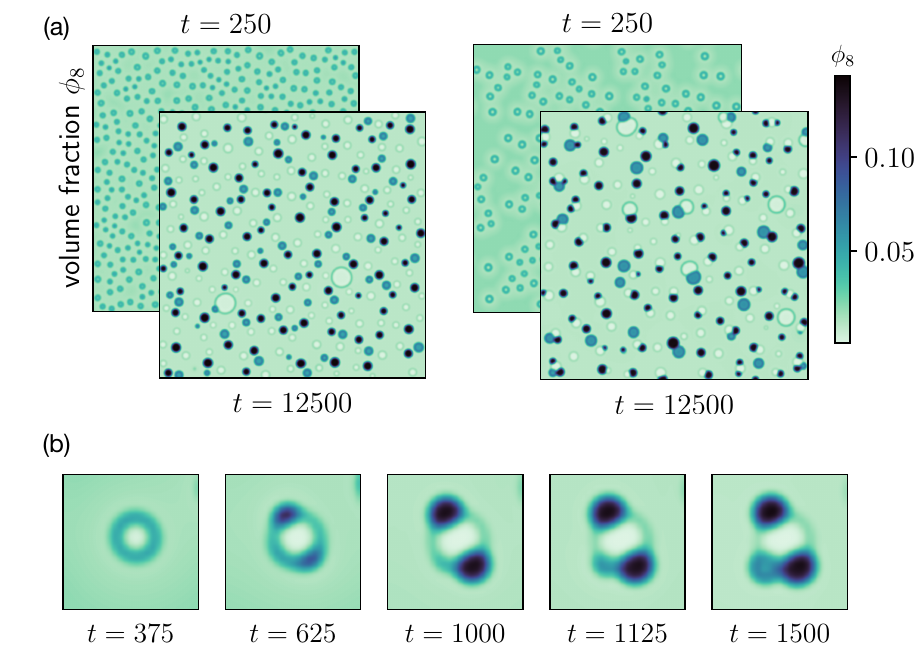}
\caption{\textbf{(a)} Dynamical evolution of a multi-component ($M=15$) mixture with the same parameter values as in previous figure, but initialized with $N(0) = 500$ (left) and $N(0) = 200$ (right) initial droplets. On the left, droplets ripen while remaining separated in space as shown in Fig.~\ref{fig:ripe_comps}. On the right, droplets grow until different phases nucleate at their interface, leading to distinct phases wetting on each other. \textbf{(b)} volume fraction $\phi_8$ snapshots}
    \label{fig:wetting}
\end{figure}

\vspace{0.5em}
\textit{Mestastable states with different wetting angles.--- }
Finally, we show that multicomponent phase-separating mixture exhibits one of the hallmarks of non-equilibrium and glassy systems, i.e. the presence of very long lived mestable states. In Fig.~\ref{fig:wetting} (a), we compare the previous simulation, with $N(0)=500$ initial droplets (left panel), with one initialized with $N(0)=200$ (right panel), leaving the other control parameters unchanged. As outlined before, in the case of $N(0)=500$, droplets differentiate into different phases. In contrast, seeding $N(0)=200$, droplets grow bigger until they become unstable and new phases nucleate at their interface. This is exemplified with a series of snapshots in Fig.~\ref{fig:wetting} (a), see also  Movie 5 in \cite{SupplementaryInformation}. This implies that, initialising two mixtures with identical control parameters, differing only in the initial spatial profiles, leads to different long-lived metastable states, one in which distinct phases are separated and one in which they wet onto each other. We believe that  the true stationary state is the one with non-zero wetting angle, since at the very end of the simulation with $N(0)=500$ droplets, two separated phases get closer and merge, see Fig.s~\ref{fig:ripe_all} and~\ref{fig:ripe_comps}, and Movie 1 in \cite{SupplementaryInformation}. Note that different morphologies can be achieved seeding different number of droplets in non-equilibrium mixtures where active chemical reactions occur, as shown in~\cite{bauermann2026Critical}.       

\textit{Conclusions.--- }
In this letter, we perform extensive numerical simulations in order to investigate the dynamics of multi-component phase-separating mixtures in different regimes. We follow the dynamics from a well-mixed phase to a metastable two-phase coexistence, which finally leads to multiple distinct phases. We found substantial changes in both dynamical exponents and the onset of ripening in comparison with binary mixtures. Simple approximations applicable to binary mixtures are expected to fail for multi-component mixtures in the initial stages of ripening. Multi-component mixtures get trapped in locally metastable states, which prevent them from ripening according to the binary theories, a phenomenon typical of phase-separation for system out of equilibrium \cite{cates2025active,li2020non}. Despite analytical progress on the dynamics of multi-component mixtures, results beyond simple linear stability are hard to get. We expect that, in the future, application of theoretical tools originally developed in the spin-glass context, such as dynamical mean-field theories \cite{garcia2024,altieri2020dynamical} and mode-coupling theories \cite{janssen2018mode,ciarella2021multi} will elucidate the richness of these dynamical regimes.

\textit{Acknowledgements.--- } 
G.B. thanks the Agencia Estatal de Investigación for funding through the Juan de la Cierva postdoctoral programme JDC2023-051554-I. 
F.O. has received funding from the Marie Skłodowska-Curie grant agreement No. 101034413. 
The authors thanks Yuting Li, Edouard Hannezo,  Jonathan Bauermann and Christoph Weber for helpful discussion.

\newpage 
\begin{appendices}

\end{appendices}





\bibliography{glassy}

\end{document}